\documentclass[aip,jcp,reprint,amsmath,amssymb]{revtex4-1}
\usepackage{graphicx}
\usepackage{dcolumn}
\usepackage{bm}
\usepackage[utf8]{inputenc}
\usepackage[T1]{fontenc}
\usepackage{mathptmx}
\usepackage{etoolbox,float}
\usepackage{booktabs}

\makeatletter
\def\@email#1#2{%
 \endgroup
 \patchcmd{\titleblock@produce}
  {\frontmatter@RRAPformat}
  {\frontmatter@RRAPformat{\produce@RRAP{*#1\href{mailto:#2}{#2}}}\frontmatter@RRAPformat}
  {}{}
}%
\makeatother

\begin{document}
\preprint{}

\title{Transitions between positive and negative charge states of dangling bonds on a halogenated Si(100) surface}

\author{T. V. Pavlova}
 \email{pavlova@kapella.gpi.ru}
\affiliation{Prokhorov General Physics Institute of the Russian Academy of Sciences, Vavilov str. 38, 119991 Moscow, Russia}

\author{V. M. Shevlyuga}
\affiliation{Prokhorov General Physics Institute of the Russian Academy of Sciences, Vavilov str. 38, 119991 Moscow, Russia}

\date{\today}

\begin{abstract}

Dangling bonds (DBs) are common defects in silicon that affect its electronic performance by trapping carriers at the in-gap levels. For probing the electrical properties of individual DBs, a scanning tunneling microscope (STM) is an effective instrument. Here we study transitions between charge states of a single DB on chlorinated and brominated Si(100)-2$\times$1 surfaces in an STM. We observed transitions between positively and negatively charged states of the DB, without the participation of the neutral state. We demonstrated that the $(+/-)$ transition occurs when the DB and substrate states are out of equilibrium. This transition is related to the charge neutrality level (CNL), which indicates a change in the DB's character from donor-like to acceptor-like. The STM voltage at which the $(+/-)$ transition took place varied depending to the electrostatic environment of the DB. Our results complement the understanding of the electronic properties of the DBs, and they should be taken into account in applications that use charge manipulation on the DBs.

\end{abstract}

\maketitle

\section{Introduction}

Dangling bonds (DBs) in silicon are amphoteric defects containing localized states in the band gap. The DB has one unpaired electron in the neutral state (DB$^0$) and it can be positively charged (DB$^+$) if the electron is removed or negatively charged (DB$^-$) if another electron is added. A scanning tunneling microscope (STM) is the most appropriate tool for probing the charge density of individual DB with atomic resolution. Several STM studies have demonstrated switching between charge states of DBs \cite{2013Bellec, 2015Labidi, 2016Kawai} as well as interesting physical phenomena such as negative magnetoresistance \cite{1989Lyo, 2016Rashidi}. Under specific scanning conditions, the charge state of the DB is determined by non-equilibrium dynamics \cite{2008Berthe, 2013Schofield, 2014Taucer}. In particular, when the STM tip is placed close to the DB, the current through the DB localized state exhibits characteristics of the current through a single-electron transistor \cite{2008Berthe, 2013Schofield, 2014Taucer}.

In addition to the fundamental interest, DBs are important from a practical point of view since they have an impact on the electronic properties of semiconductor devices by capturing carriers in localized states. On the Si(100) surface, DBs are used for the incorporation of impurities \cite{2003Schofield}, and, in addition, there are proposals for their application in nanoelectronics \cite{2018Huff, 2017Yengui, 2024Pitters} and for modeling two-dimensional structures \cite{2013Schofield, 2018Wyrick}.

In the majority of STM studies, localized states of the DB were considered on the Si(100)-2$\times$1-H surface \cite{2018Huff, 2013Schofield, 2018Wyrick, 2013Bellec, 2015Labidi, 2016Kawai, 2016Rashidi, 2014Taucer}. In this work, we chose a halogenated (Cl, Br) Si(100) surface to study DBs. An additional motivation for this work is the recent interest in the application of DBs on a halogenated surface for positioning impurities with atomic accuracy \cite{2018Pavlova, 2019Dwyer, 2020Pavlova, 2022Pavlova, 2021Pavlova}. It has already been shown on Si(100)-Cl and -Br surfaces that single DBs can be created and their charge states can be manipulated in an STM \cite{2022Pavlova}. It has also been demonstrated that a change of the DB charge affects the reactivity of the Si(100)-Cl surface toward PBr$_3$ molecules \cite{2024Pavlova}.

Here, we studied switching between different charge states of DBs on a halogenated (Cl, Br) Si(100) surface. We have demonstrated that charged and uncharged DBs behave differently. For charged DBs, we observed a $(+/-)$ transition that occurs out of equilibrium between the DB and substrate levels. In addition to a single DB on Si(100) in a halogen vacancy, we investigated a single DB located in the third Si layer on the chlorinated Si(100) surface.

\section{Experimental and computational details}

The investigations were carried out in an ultra-high vacuum (UHV) with a base pressure of 5$\times$10$^{-11}$\,Torr.  All measurements were performed using a low-temperature scanning tunneling microscope GPI CRYO (SigmaScan Ltd.) operated at 77\,K. We used both B-doped (p-type, 1\,$\Omega$\,cm) and P-doped (n-type, 0.1\,$\Omega$\,cm) Si(100) samples. Samples were prepared by outgassing at 870\,K for several days in UHV, followed by flash annealing at 1470\,K. After that, Cl$_2$ or Br$_2$ was introduced with a partial pressure of 10$^{-8}$\,Torr during 100--200\,s at a sample temperature of 370--420\,K. Both electrochemically etched polycrystalline W and mechanically cut Pt-Rh STM tips were utilized. All STM images were recorded at a positive bias voltage on the sample. The WSXM program \cite{WSXM} was used to process STM images.

The spin-polarized DFT calculations were carried out using the Vienna \textit{ab initio} simulation package (VASP) \cite{1993Kresse, 1996Kresse} within the framework of generalized gradient approximation (GGA) of Perdew-Burke-Ernzerhof (PBE) for the exchange-correlation potential \cite{1996Perdew}. The ion-electron interaction was modeled by the projector-augmented wave method \cite{1999Kresse, 1994Blochl} with a kinetic-energy cutoff of 400\,eV.
To include the van der Waals correction, the DFT-D2 method developed by Grimme \cite{2006Grimme} was utilized. The Si(100)-2$\times$1 surface was modeled using periodic 6$\times$6 supercells. A slab model consisting of sixteen Si layers was used. To avoid the influence of surface-surface interactions, we created a vacuum space between slabs of 15\,{\AA} thickness. Chlorine (bromine) atoms were placed on the upper surface to form a Si(100)-2$\times$1-Cl (-Br) structure, while the bottom Si(100) surface was passivated with hydrogens. The bottom two Si layers were set during the optimization process in their bulk positions, while the coordinates of other atoms were fully relaxed. The optimal structure was obtained when the residue forces acting on the relaxed atoms were less than 0.01\,eV/\,{\AA}. To simulate positive or negative charge states, the electron was removed or added to the supercell, respectively. The 4$\times$4$\times$1 $\Gamma$-centered k-point mesh was used. STM images were calculated within the Tersoff-Hamann approximation \cite{1985Tersoff}.

\section{Results and Discussion}

\subsection{DB on the halogenated Si(100)-2$\times$1 surface}

\begin{figure}[h]
 \begin{center}
 \includegraphics[width=\linewidth]{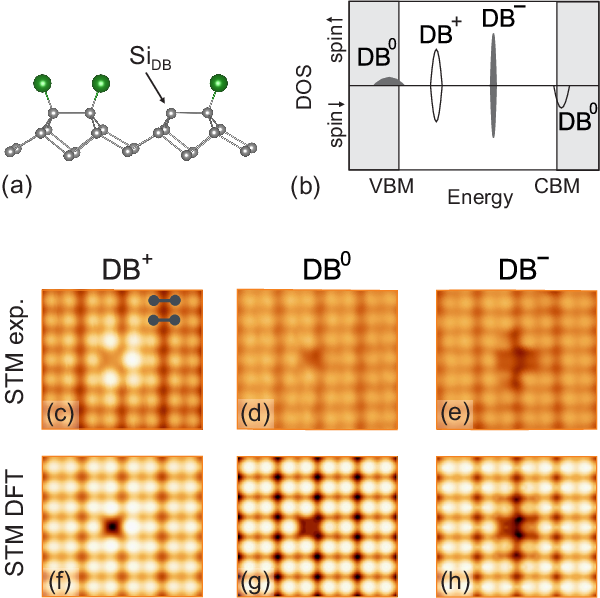}
\caption{\label{fig_DOS_STM} (a) Structure of the Si(100)-2$\times$1-Cl surface with a Cl vacancy. Si atoms are shown in gray and Cl in green. The silicon atom holding the DB is denoted as Si$_{DB}$. (b) Schematic representation of spin-resolved DOS of the Si atom holding the DB$^+$, DB$^0$, or DB$^-$. Filled levels of the DB are shaded. Experimental (c--e) and simulated (f--h) empty state STM images of the DB$^+$, DB$^0$, and DB$^-$ on the Si(100)-2$\times$1-Cl surface. Experimental images were obtained at 3.9\,V (c), 3.6\,V (d), and 1.8\,V (e). Theoretical images were calculated at 2.5\,V (f), 2.3\,V (g), and 2.0\,V (h). Silicon dimers are marked by dumbbells in (c).}
\end{center}
\end{figure}

On the Si(100)-2$\times$1 surface covered with a halogen monolayer, Si atoms form three bonds with nearby silicon atoms and one bond with a halogen atom. The silicon atom without the halogen atom (Si$_{DB}$) holds a DB (Fig.~\ref{fig_DOS_STM}a). The DB levels are illustrated schematically in Fig.~\ref{fig_DOS_STM}b based on the DFT calculations of Ref.~\cite{2022Pavlova}. A neutral dangling bond (DB$^0$) is unsaturated and has one electron occupying a level around the valence-band maximum (VBM). If this electron is removed, the DB becomes positively charged (DB$^+$) and has a doubly unoccupied level inside the band gap. If an electron is introduced to the unoccupied level of the DB$^0$, which is near the conduction-band minimum (CBM), the DB becomes negatively charged (DB$^-$). The DB$^-$ doubly occupied level lies in the middle of the band gap, above the DB$^+$ level. A change in the DB charge state alters the geometry of the surface structure. Specifically, different numbers of electrons on the Si$_{DB}$ result in different sp-hybridization with the states of the nearest three Si atoms, which affects the tilt angle of the Si dimer holding the DB on a halogenated Si(100)-2$\times$1 surface \cite{2022Pavlova}.

STM images of the Si(100)-2$\times$1-Cl surface with a DB in three different charge states are shown in Figs.~\ref{fig_DOS_STM}c--h. When the DB is neutral, the STM image simply displays the Cl vacancy (Fig.~\ref{fig_DOS_STM}d). Unoccupied states of the DB$^+$  appear as a bright halo in the empty state STM image, extended to the adjacent Cl atoms (Fig.~\ref{fig_DOS_STM}c). In contrast, the DB$^-$ with doubly occupied states is visualized as a dark halo in the empty state STM image (Fig.~\ref{fig_DOS_STM}e). Note that the simulated STM images (Figs.~\ref{fig_DOS_STM}f--h) agree well with the experimental ones (Figs.~\ref{fig_DOS_STM}c--e). In this work, we investigate DBs on chlorinated and brominated silicon surfaces. Despite the different electronegativity of these halogens, previous calculations reveal that the DB levels \cite{2022Pavlova} and STM images \cite{2022PavlovaJChemPhys} are very similar on both surfaces.

\begin{figure}[h]
 \begin{center}
 \includegraphics[width=0.7\linewidth]{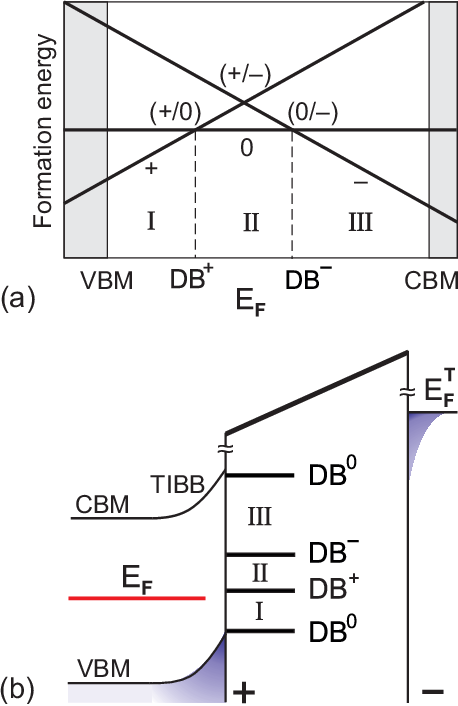}
\caption{\label{fig_diagram} (a) Schematic representation of the formation energy as a function of the Fermi energy for different charge states of the DB on the halogenated Si(100)-2$\times$1 surface. The DB$^+$, DB$^0$, and DB$^-$ are in equilibrium with the sample if the Fermi level of the sample (E$_F$) is located in ranges I, II, or III, respectively. (b) Energy band diagram for empty-state STM imaging of the DB on the halogenated Si(100)-2$\times$1 surface. In three different charge states, the DB has different levels. A positive voltage applied to the sample induces the upward band bending, which can be used to change the position of the DB levels relative to the E$_F$. E$^T_F$ is the Fermi level of the tip.}
\end{center}
\end{figure}

To plot the formation energy as a function of the Fermi energy for DBs on the halogenated Si(100) surface (Fig.~\ref{fig_diagram}a), we use the analogy with bulk DBs \cite{2024Varley}. According to the charge state stability diagram for bulk Si DBs, positive, neutral, or negative DB states are stable depending on the position of the Fermi level. The three lines in the diagram correspond to the three charge states of the DB, and their intersections denote the charge transition levels, $(+/0)$, $(0/-)$, and $(+/-)$. In the diagram, level $(+/0)$ is located below level $(0/-)$, indicating that the DB in bulk silicon behaves as a positive-U defect (for negative-U defects, level $(+/0)$ is located above $(0/-)$) \cite{2024Varley}. The formation energy on the surface and in the bulk may differ, since lattice relaxation plays a major role in formation energy. In particular, lattice relaxation can be different for charge states with different orbital hybridizations and thus can change the order of transition levels. For a hydrogenated Si(100) surface, the Coulomb energy U is 0.34 eV \cite{2013Schofield}. For a halogenated Si(100) surface, U has almost the same value since U is calculated as the difference in energy of the DB levels, which are similar on the hydrogenated and halogenated surfaces \cite{2022Pavlova}. Therefore, U values for DBs in the bulk and on the surface are positive. In STM experiments, positive U on the halogenated surface is confirmed by the existence of DBs in all three charge states (Figs.~\ref{fig_DOS_STM}c--e) which are stable on the diagram (Fig.~\ref{fig_diagram}a). Thus, the diagrams for DB for both bulk silicon and the surface are qualitatively similar that allows us to use the diagram for bulk Si DBs. Note that to explain the DB charge transitions, we use only the qualitative diagram, without quantitative values of the transition levels.

In thermodynamic equilibrium of the DB with the substrate, all three charge states should be observed depending on the position of the Fermi level in the bandgap (Fig.~\ref{fig_diagram}a). If the Fermi level is in range I, the unfilled level of the DB$^+$ is above the Fermi level, and the DB$^+$ is in equilibrium. In range II, the DB$^0$ is in equilibrium, and it has filled level below and unfilled level above the Fermi level. In range III, the DB$^-$ is in equilibrium, and its filled level is below the Fermi level.

To activate the charge state transitions of the DB, the effect of tip-induced band bending (TIBB) can be used \cite{2013Bellec, 2015Labidi}. Bias voltage in the STM causes TIBB in the sample, and thereby affects the positions of the DB levels relative to the Fermi level (Fig.~\ref{fig_diagram}b). In Fig.~\ref{fig_diagram}b, we placed the Fermi level approximately in the middle of the band gap, because the samples used in our experiments are more intrinsic in the near-surface region of about tens of nanometers due flash annealing, which resulted in dopants depletion near the surface \cite{2012Pitters}. In equilibrium, TIBB should allow for sequential switching of the DB's charge states. The charge stability diagram (Fig.~\ref{fig_diagram}b) shows that reducing the voltage on the sample from high to low leads to a decrease in the band bending. As a result of the decrease in band bending, the Fermi level falls sequentially into ranges I, II and III. This can cause a change in the DB charge state from positive to neural and then to negative, respectively (Fig.~\ref{fig_diagram}a).

\subsection{STM experiments}

\begin{figure}[h]
 \begin{center}
 \includegraphics[width=0.8\linewidth]{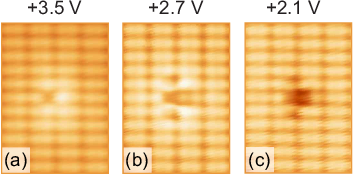}
\caption{\label{Fig_vac_ad1} Changing of the DB charge state from positive to negative by decreasing the positive bias voltage. (a--c) Empty state STM images of the Si(100)-2$\times$1-Cl surface with the DB (P-doped Si). STM images are independent of whether the voltage goes from high to low or vice versa. Note that doubling the current (from 1 to 2 nA) when recording an STM image at 2.7\,V (b) did not result in a change in the DB visualization.}
\end{center}
\end{figure}

Figure~\ref{Fig_vac_ad1} shows the switching of the DB charge state on the Si(100)-2$\times$1-Cl surface as the positive bias voltage decreases. At a high voltage, the TIBB is strong and hence the Fermi level is located in range I in Fig.~\ref{fig_diagram}b. At high voltage, the DB is positively charged (Fig.~\ref{Fig_vac_ad1}a), which is indicated by the characteristic protrusion surrounding it (Fig.~\ref{fig_DOS_STM}c). At a low voltage, the TIBB is small, so the Fermi level is located in range III. The DB is negatively charged at a low voltage as evidenced by a dark halo surrounding it (Fig.~\ref{Fig_vac_ad1}c) in accordance with the STM image of the DB$^-$  (Fig.~\ref{fig_DOS_STM}e). In the intermediate voltage range between high and low voltages (range II), the STM image of the DB (Fig.~\ref{Fig_vac_ad1}b) does not agree with the STM images for either DB$^0$, DB$^+$, or DB$^-$ (Figs.~\ref{fig_DOS_STM}c--e). In particular, the STM images of the DB in range II (Fig.~\ref{Fig_vac_ad1}b) do not agree with the STM image of the DB$^0$ because the DB$^0$ lacks a halo (Fig.~\ref{fig_DOS_STM}d). Thus, we conclude that in range II, DB is not in equilibrium with the substrate, since it shows an unusual visualization instead of a neutral state.

\begin{figure*}[t]
 \begin{center}
 \includegraphics[width=\linewidth]{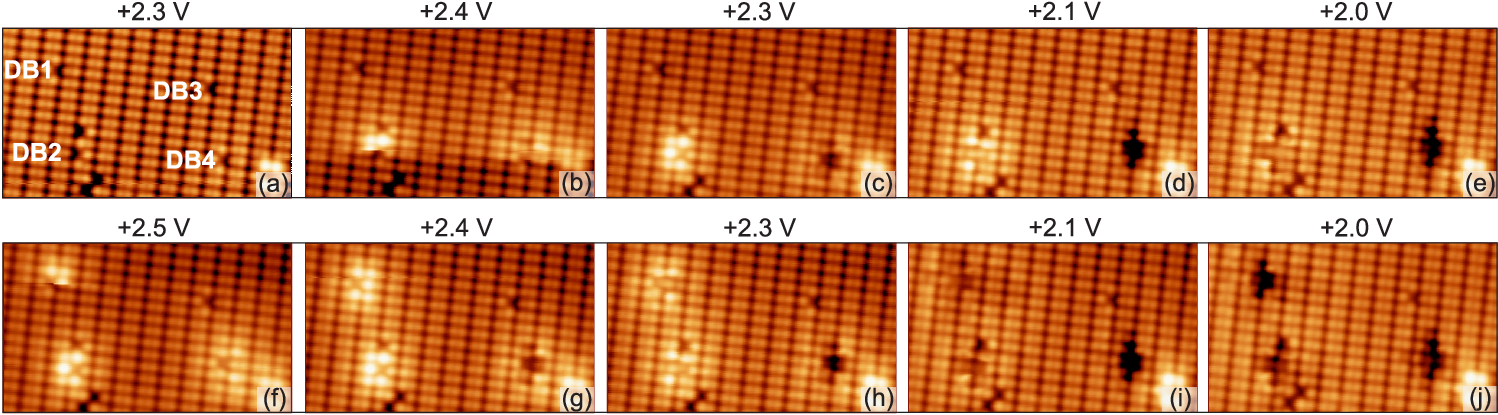}
\caption{\label{Fig_DB0} Changing of the DBs charge state by varying the sample voltage. Empty state STM images (10.0$\times$5.6\,nm$^2$, I$_t$ = 2\,nA, B-doped Si) of the Si(100)-2$\times$1-Cl surface with several Cl vacancies.  Initially, all DBs were neutral (a). When DB2 and DB4 were charged (b), their charge states changed with voltage variation, and the unusual visualization was observed (b--j). The voltage variations have no effect on the neutral vacancies with DB1 and DB3 (a--e). When an electron was removed from DB1 (f), DB1 began to change its state when the voltage was changed (g--j). The slow scan direction proceeded from bottom to top.}
\end{center}
\end{figure*}

Remarkably, we have observed the unusual visualization of the DB as in Fig.~\ref{Fig_vac_ad1}b only when scanning the initially charged DBs. Figure~\ref{Fig_DB0} shows sequentially recorded STM images of the Si(100)-2$\times$1-Cl surface with several vacancies at various voltages. Initially, all DBs were neutral when scanned at 2.3\,V (Fig.~\ref{Fig_DB0}a). When the voltage was increased to 2.4\,V (Fig.~\ref{Fig_DB0}b), DB2 and DB4 were charged due to the upward TIBB. Note that due to nonidentical local electrostatic potential caused by adjacent impurities and defects, the voltage at which the DB changes its charge state can differ between DBs \cite{2015Labidi, 2022Pavlova}. After charging two DBs, we reduced the voltage (Figs.~\ref{Fig_DB0}c--e) to pass through the range II. The unusual visualization of DB2 and DB4 was observed at 2.0\,V and 2.3\,V, respectively. However, neutral DBs did not change their charge state (Figs.~\ref{Fig_DB0}c--e). To charge another neutral DB, we used a higher voltage (2.5\,V) than when charging DB2 and DB4 (2.4\,V). As a result, DB1 was charged (Fig.~\ref{Fig_DB0}f). We then lowered the voltage again to pass through the range II (Figs.~\ref{Fig_DB0}g--j). Besides DB2 and DB4, DB1 now showed the unusual visualization at 2.3\,V, while DB3, which remained neutral, did not modify its charge state. Thus, a change in voltage causes transitions between positively and negatively charged states, while the neutral DB remains uncharged.

\begin{figure}[h]
 \begin{center}
 \includegraphics[width=\linewidth]{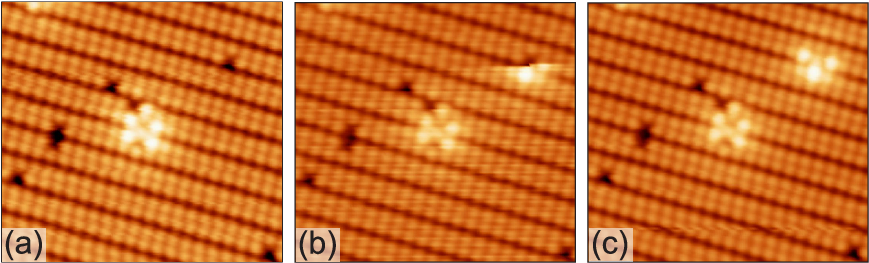}
\caption{\label{Fig_spont} Spontaneous transition from the DB$^0$ to DB$^+$ on the Si(100)-2$\times$1-Cl surface. Empty state STM images (7.4$\times$6.8\,nm$^2$, $U_s =+3.8$\,V, I$_t$ = 1.0\,nA, B-doped Si) were sequentially recorded. The DB$^0$ in the upper right corner looks dark (a), but after an electron removal (b), it becomes positively charged and looks bright (c) as the DB$^+$. In (b), the slow scan direction proceeded from top to bottom.}
\end{center}
\end{figure}

Along with the voltage-controlled change of the DB charge state, the DB may also spontaneously charge while being scanned, however the probability of this process is very rare. Figure~\ref{Fig_spont} shows an example of an electron removing from the neutral DB. Initially, all of the DBs were neutral (Fig.~\ref{Fig_spont}a), but then one DB$^0$  switched to the positive charge state (Fig.~\ref{Fig_spont}b), and, upon further scanning, the charge state of all DBs remained unchanged (Fig.~\ref{Fig_spont}c). We also very rarely observed the reverse process of spontaneous discharge of the DB during scanning, i.e., the transition from a charged state to a neutral one. After being discharged, the DB behaved like the DBs$^0$ in Fig.~\ref{Fig_DB0} and no longer exhibited the unusual visualization like in Fig.~\ref{Fig_vac_ad1}b.

\subsection{Discussion}

To switch the charge states of DBs, we varied the voltage in our STM experiments. We observed DBs$^+$ at high voltage, DBs$^-$ at low voltage, and the unusual visualization of DBs (Fig.~\ref{Fig_vac_ad1}b) at the intermediate voltages (range II). In range II the unusual visualization of the DB does not agree with the image of the DB$^0$, therefore switching between the DB$^+$ and DB$^-$ does not occur through the neutral state of the DB. Furthermore, we observed that the unusual visualization of the DB in range II disappears when the DB discharges spontaneously. Thus, the DB$^0$ is not realized in voltage range II, despite having the lowest energy. As a result, we were unable to observe the thermodynamic transition levels $(0/+)$ and $(0/-)$ because the charge state did not relax to its equilibrium state DB$^0$. Consequently, in our experiment, a nonequilibrium situation is realized, and we observed the $(+/-)$ transition, which has a higher energy than the DB$^0$ in range II (Fig.~\ref{fig_diagram}a). The $(+/-)$ transition occurs between charged states without any involvement of a neutral state; therefore, a two-electron process is realized.

The $(+/-)$ transition appears in the absence of equilibrium between the substrate and DB levels, consequently the filling rate of the DB level is not equal to its emptying rate. The dark area in the center of the DB in the empty state STM image (Fig.~\ref{Fig_vac_ad1}b) indicates that the state is filled, therefore the DB in negatively charged. Indeed, DB$^-$ has a doubly occupied state in the band gap and is visualized as a dark area in the empty state STM image (Fig.~\ref{fig_DOS_STM}e). Because the DB is negatively charged, it is most of the time loaded with electrons, hence its filling rate exceeds its emptying rate. In the STM images in range II (Fig.~\ref{Fig_vac_ad1}b), the dark area is surrounded by a bright halo, indicating a positive charge of the DB when the tip is not yet above the DB. Indeed, unoccupied DB$^+$ states result in a bright halo in an empty state STM image (Fig.~\ref{fig_DOS_STM}c). Thus, in range II, the DB is negatively charged when the tip is directly above it and positively charged when the tip is a short distance away from it.

\begin{figure}[h]
 \begin{center}
 \includegraphics[width=\linewidth]{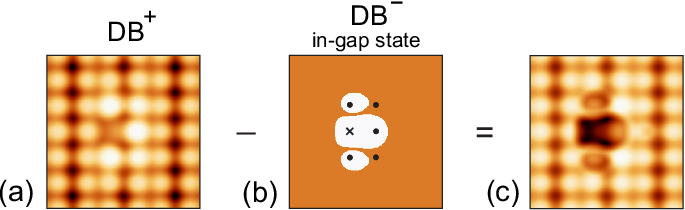}
\caption{\label{Fig_DFT_STM} DFT modeling of the STM images of the $(+/-)$ transition of the DB. (a) Calculated STM image of the DB$^+$ at 1.5\,V, (b) calculated STM image of the DB$^-$ integrated inside the band gap, and (c) simulated STM image of the DB with a dark area in the center, obtained as the difference between the images (a) and (b). In the calculated STM image of the in-gap states (b), the positions of Cl are marked with dots, and the Cl vacancy is marked with a cross.}
\end{center}
\end{figure}

To simulate the STM image of the $(+/-)$ transition, we took into account that the DB has different charge states depending on the tip position. When the tip is far from the DB, the DB is positively charged, so we calculated an STM image from the DB$^+$ (Fig.~\ref{Fig_DFT_STM}a). When the tip is close to the DB, we calculated an STM image from the DB$^-$ state located in the band gap (Fig.~\ref{Fig_DFT_STM}b), which was obtained by integrating the charge density over the energy region of this state. The DB$^-$ in-gap state has a three-lobed shape in the STM image, which is explained by hybridization with states of surrounding atoms (the 3D distribution of electron density is given in SM). To simulate the final STM image, we superimposed the STM image from the DB$^-$ in-gap state atop the STM image of the DB$^+$ (Fig.~\ref{Fig_DFT_STM}c). The DB$^-$ state is filled, so we used subtraction instead of addition because we modeled the STM image of an empty state. As a result, we were able to reproduce the experimental STM image of the DB in the case of the $(+/-)$ transition (Fig.~\ref{Fig_vac_ad1}b). The calculated spatial distribution of the electron density of the DB$^-$ in-gap state (Fig.~\ref{Fig_DFT_STM}c) appears to be very similar to the three-lobed depression inside the DB (Fig.~\ref{Fig_vac_ad1}b).

The energy of the $(+/-)$ transition is consistent with the so-called charge neutrality level (CNL) that used as a reference point for band alignment in semiconductors \cite{2003Walle, 2024Varley}. Near the CNL, the donor-acceptor character of the DB states changes. Indeed, at the $(+/-)$ transition, the Fermi energy is equal to the energy of the DB$^+$ and DB$^-$ levels. Usually, the CNL can be measured experimentally as the average value for the whole sample. In our STM experiments, we observed the $(+/-)$ transitions for individual DBs. In fact, we identified local CNL for an individual DB since the non-uniform electrostatic potential of the surface affects the energy levels for each given DB.

The $(+/-)$ transitions are not a unique feature of an isolated DB at a Si atom on top of the chlorinated Si(100)-2$\times$1 surface. We also observed the $(+/-)$ transitions for single DBs at a third-layer silicon atom and for DBs on a brominated Si(100)-2$\times$1 surface. The $(+/-)$ transitions in these cases are very similar to those described above for the DB in a Cl vacancy, except that DB in a bromine vacancy can hop within the Si dimer under certain scanning parameters \cite{2022PavlovaJChemPhys}. STM images and discussion for these cases are presented in SM.

We believe that the $(+/-)$ transitions can also occur on DBs on a hydrogenated Si(100) surface in a non-equilibrium regime,  although halogens are more electronegative compared to hydrogen. The higher electronegativity of chlorine causes a higher polarization of the Si-Cl bond compared to Si-H. As a result, the work function of the Si(100)-2$\times$1-Cl surface is about 1.0 eV higher than that for Si(100)-2$\times$1-H, according to our calculations. As was shown for the Si(111)-1$\times$1-Cl surface \cite{2006Blomquist}, the shift of the electrostatic potential from the vacuum level does not lead to the appearance of band bending and surface states in the band gap. In the case of Si(100)-2$\times$1-Cl, the surface states also lie outside the band gap \cite{1998Lyubinetsky, 2006Nakayama}. Therefore, electronegative chlorine does not induce band banding, leaving the bands flat, as in the case of hydrogen on the Si(100) surface. According to the previous calculations \cite{2022Pavlova}, the DB levels on the Si(100) surface covered with chlorine, bromine and hydrogen are very similar. In addition, although the electronegativities of Cl and Br are different, we found no difference in experiments carried out under the same conditions on chlorinated and brominated silicon surfaces. Thus, we believe that the two-electron $(+/-)$ transitions of the DBs can be observed on the Si(100) surfaces covered by different adsorbats.

	On the hydrogenated Si(100)-2$\times$1 surface, DBs were studied in the regime of non-equilibrium dynamics  \cite{2013Schofield, 2014Taucer, 2011Livadaru}. In Ref.~\cite{2014Taucer} it was supposed that transitions between charge states of the DB include the neutral state, however, the experiments were carried out in a different geometry, when the tip was located far from the DB. Remarkably, no sequential switching from positive to negative through neutral state was demonstrated (such that STM images for the three stable states were recorded), which would contradict our findings on halogenated surfaces.

\section{Conclusions}

The charge switching on DBs on the halogenated Si(100) surface was investigated in STM. When manipulating the DB charge by lowering the positive voltage on the sample, the charge changes from positive to negative, but the neutral DB is not involved in the charge transition. We demonstrated that this effect can be observed only if the DB is initially charged. As a result, we observed the two-electron $(+/-)$ transition that does not occur at equilibrium. The $(+/-)$ transition, at which the donor-acceptor character of DB states changes, is related to the concept of CNL. The STM images at the $(+/-)$ transition were simulated by superimposing calculated STM images of the positively and negatively charged DB. We have also demonstrated that non-equilibrium charging can be observed in STM on the DBs in the third silicon layer and on the DBs on a brominated surface. The obtained results complement the understanding of non-equilibrium phenomena that occur when current flows through the DB. The absence of the neutral state in the transitions between charged states of the DB should be taken into account when manipulating the DB charge.

\section*{Supplementary Material}
See the supplementary material for density of states for charged DBs on Si(100)-2$\times$1-Cl and the $(+/-)$ transition of DBs in the third Si layer below the surface and on the brominated Si(100)-2$\times$1 surface.

\begin{acknowledgments}
This study was supported by the Russian Science Foundation under grant No. 21-12-00299. We also thank the Joint Supercomputer Center of RAS for providing the computing power.
\end{acknowledgments}

\bibliography{paper_DB_2024_PCCP_rev_arxiv}
\bibliographystyle{rsc}

\end{document}